\documentclass[twocolumn,showpacs,preprintnumbers,amsmath,amssymb]{revtex4}

\usepackage{graphicx}
\usepackage{dcolumn}
\usepackage{bm}

\begin{document}

\preprint{}

\title{AN ALGORITHMIC INFORMATION THEORY APPROACH TO THE EMERGENCE OF ORDER USING SIMPLE REPLICATION MODELS}

\author{Sean D Devine}
 \altaffiliation {Victoria Management School, Victoria University of Wellington, PO Box 600,
Wellington, 6140, New Zealand}
\email {sean.devine@vuw.ac.nz}

\date{\today}

\begin{abstract}
Algorithmic entropy is closely related to the entropy measures of Shannon, Boltzmann and Gibbs.  This paper applies the algorithmic entropy concept to simple examples of replication processes to illustrate how replicating structures can generate and maintain order in a non equilibrium system.  Variation in replicating structures enhances the system's ability to maintain homeostasis in a changing environment by allowing it to evolve to a more restricted region of its state space. Stability is further enhanced when replicating systems develop dependencies, by sharing information or resources.  Such systems co-evolve, becoming more independent of the external environment. Nested systems have a hierarchy of dependency but have low algorithmic entropy as they are in principle simpler to describe algorithmically. Nesting of replicating systems offsets the algorithmic entropy increase that occurs through the variation in the replicated substructures, by allowing the system itself to increase in organizational ordering with little increase in algorithmic entropy.  Chaitin's d-diameter complexity provides a measure of the level of order in nested replicating systems.  

\it {keywords} \rm Structures and organization in complex systems; dynamics of evolution; replication; algorithmic information theory; algorithmic entropy.  
\end{abstract}

\maketitle
\section{Introduction}

While it is understood that the universe started in a highly ordered state, it is not clear what mechanisms underpin the local, far-from-equilibrium order that has emerged as the universe trends towards equilibrium.  This paper, by using simple models as illustrations, argues that much of the order observed is analogous to a crystallization process obeying physical laws. I.e. the physical process of replication accesses existing order, usually in the form of high grade energy, and ejects disorder to create new ordered structures. Algorithmic Information Theory (AIT) provides a framework to look at the emergence of order through replication processes and to track the corresponding entropy changes.   This framework utilises the fact that the simplest algorithmic description of a structure or system is an entropy measure that is closely related to the Shannon and Boltzmann-Gibbs understandings of entropy.  The application to simple models provides insights into the emergence of order giving confidence that the approach is applicable in principle to far more complex ordered systems.  While the models are illustrative, the following general principles emerge. 
\begin{itemize}

\item Structures generated by replication processes are highly ordered, having low algorithmic entropy, and are more likely to emerge than similar structures produced by random fluctuations or non replicating physical processes.
 \item 
Replication processes can maintain an ordered system away from equilibrium in an attractor-like region of the system's dynamical state space.
 \item Where a system generates similar, but not identical replicated structures, the algorithmic entropy increases.  However, variability in the replication allows the dynamical system to maintain homeostasis in a changing environment by providing a mechanism for the system to evolve to a more restricted region of its state space - i.e. diversity in the replication process can maintain the system in a stable configuration through adaptive evolutionary-like  processes. 
\item
Homeostasis is characterised by a set of states having the same provisional algorithmic entropy.
 \item 
Coupled replicator systems create greater stability against change by co evolving.
\end{itemize}

\subsection{Replicators}

A replicator, such as an autocatalytic set, or a bacterium in an environment of nutrients, is a physical or biological structure that is able to reproduce itself by utilizing energy and resources from an external environment. Similar replication occurs where a crystal forms from a melt or where spins align in a magnetic material. 

In a resource rich environment, where the probability that a structure will replicate increases with the number $n$ of existing structures, replicated structures are more likely to be observed than alternative structures. I.e. if the probability is proportional to \(n^x\) (where $x>0$), d$n$/d$t$ = $kn^x$ \cite {1.}.  For example, molecules are more likely to solidify on a seed crystal in a melt and, given one strand of DNA in the right environment, the probability of a second strand of DNA appearing is comparatively high. In this paper, the word `replicate' will be used to distinguish the copied structure, such as the DNA information string, from the full replicating system which, in the DNA case, is the whole cell.

Where resources are limited, the number of replicates grows over time until a state of homeostasis is reached; a state where the set of replicates and the environment reach a long-term stable relationship.  The somewhat simplistic logistic growth equation,

\begin{equation}
    {\rm{d}} n/{\rm d} t = (n/\tau)(1   - n/k)  				   
\end{equation}   
captures the essence of replication. Here $\tau$ represents the usual exponential growth time and  $k$ represents the carrying capacity of the system,  Replicates may die and be born but, where the birth rate = death rate of replicates (i.e. \({\rm d}n/{\rm d}t=0)\), the system is maintained in a homeostatic state by the flow through of nutrients and energy. Such a replicating, homeostatic system would appear to be the simplest conceptual description of a `far-from-equilibrium' existence.

Partial crystallization, or partial alignment of magnetic spins just above the system's transition temperature, are perhaps the simplest examples of homeostatic systems.  The ordered crystalline material or aligned spins only exist in a limited region of space but, at the boundaries of this region, replicates die and are born. However, below the transition temperature, the ordering process eventually consumes all the resource - the order is `frozen'.  Similarly an isolated bacterial system may not reach a homeostatic situation but will grow until all the resources are consumed.  

Replication may produce order through the repeats of the actual structures, such as an autocatalytic set, or a bacterium.  Alternatively spatial coherence or correlation between structures emerges through replication.  Examples are: gaseous condensation, crystallization, magnetic phase transitions or coherent photons. For these systems, latent heat is extracted and passed to the environment.  There are also more mixed systems where, for example, cells replicate and then amalgamate to form further order as specialist tissue.  In a living system, part of the nutrient environment is formed from other living systems.  The emergence of life can be viewed as replicating systems becoming interdependent to form larger ordered structures. 

Replicated structures are simpler to describe than systems of non replicated structures.  The next section shows how algorithmic information theory formalizes this simplicity by defining the algorithmic entropy or complexity of a structure in terms of its simplest description.  The approach provides a tool to quantify the order in the replicated structures.
\section{Algorithmic Information Theory}

Chaitin \cite{2.} and Kolmogorov \cite{3.} have developed algorithmic information theory (AIT) to define the complexity of a string `$s$' in terms of the length of the shortest binary algorithm $p^*$ that is able to generate that string.  The complexity is denoted by \(|p^*|\), where the vertical lines denote the length of the enclosed algorithmic description. The length of this algorithm, when defined appropriately, is also a measure of the entropy and the information content of the string (see later). However readers should note that the use of the word `complexity' in this context differs from that used in the natural sciences. In science, `complexity' generally is used to describe `pattern' or `organization'.  In the AIT framework the most complex strings are those that show no pattern and cannot be described by shorter algorithms.  

While in principle, any structure can be described by a string and its complexity measured by \(|p^*|\), $p^*$ is non computable.  There is no certainty that a compressed or short description of $s$ exists.    
The string of $N$ repeated ones; i.e. $s = 11111 \ldots $  , is highly ordered and can be described by the algorithm \(p' = PRINT \;{\rm 1 }, \; N \; times \).  If a binary algorithm for $p'$ generates $s$ on a computer $C$, the complexity \(|p^*|\) must be $\leq |p'|$; i.e.

\begin{align}
|p^*| \leq |p'| &= |N|+ the \:size \: of \: the \: code \: for \: the \nonumber\\
 &PRINT \: instruction \nonumber\\
&+ the \:number \: of \: bits \: to \: specify \:the \:object.  
\end{align}                               
For large $N$ the size of the above algorithm becomes:

\begin{equation}
|p^*| \approx |p'| = log_2N +|1| + O(1),
\end{equation}
where the $O(1)$ refers to a string that is of the order of 1, i.e. independent of $N$. On the other hand an algorithmically random string is one that exhibits no pattern. Each element of the string must be specified e.g. by:
\begin{equation}
|p'| = PRINT \; s ,
\end{equation}
and \(|p^*|\) will be greater than the length of $s$. 

In practice, to avoid ambiguity, it is usual to code the algorithm describing the string using a prefix (free), or self-delimiting code. This also has the advantage that the algorithmic entropy measure, which is based on self-delimiting coding is closely aligned with traditional entropies.  In such a formalism, no code is a prefix of any other. The algorithmic complexity of string $s$ using self-delimiting coding is commonly denoted by $K(s)$ or, where the input string $i$ is given, $K(s|i)$.

In practice self-delimiting coding adds about $log_2N$ to a string that otherwise would be of length $N$. The associated algorithmic entropy $H_{algo}$ of a string, effectively its information content, is defined as $H_{algo} \equiv K(s)$.  In what follows, algorithmic entropy rather than algorithmic complexity will be the preferred term. However, where the shortest length of an algorithmic description does not necessarily correspond to the algorithmic entropy, because there are restrictions on the type of algorithm being used, the length of the restricted algorithm will be defined by $K$ or, for a reversible computation, $KR$.

Using such coding, Chaitin \cite {4.} has shown that the computer dependence of the complexity can be mostly eliminated by defining the complexity using a reference Universal Turing Machine (UTM) that can simulate any other Turing Machine.  If  $p^*$ is the shortest programme that generates $s$ on the reference UTM, $p^*$ is within a fixed constant O(1) of any alternative.  Hence
$$H(s) \leq |p^*| + O(1).$$
The $O(1)$, which may be a few hundred bits, covers the computing overheads to specify the operation of the reference UTM.  By choosing a simple UTM, this term can be made negligible relative to the size of strings that might specify the microstates of a thermodynamic system. 

There is an ambiguity in coding a natural number $N$ in binary form as ``01'' is the same as ``1''.  This can be avoided if lexicographic ordering is used \footnote{Lexicographic ordering is used to define codes for the natural numbers unambiguously.  E.g. the integers from 0 to 6 etc are coded as  $\varnothing$ , 0, 1, 00, 01, 10, 11,  etc, where $\varnothing$  is the empty string.}. In this case, the length of $N$'s description is  $\lfloor log_2(N+1)\rfloor$.  Here the floor notation denotes the greatest integer $\leq$ to the enclosed term.  This integer is equal to, or one bit less than $\lceil log_2N\rceil$, i.e. $log_2N$ rounded up. As is the custom, in what follows $log_2N$ will be taken to mean the integer $\lceil log_2N\rceil$.

However, a self-delimiting code for $N$ must also contain implicit information about the size of $N$, but where $N$ comes from an unknown probability distribution, there is no simple process to make a representation self-delimiting.  One possibility is to use $codeN =1^{|N|}0N$, which has length $2|N| +1$; i.e. is close to $2log_2N+1$.  More sophisticated coding procedures can produce a shorter self-delimiting description of $N$ by including the length of the code for $N$  and the length of the length of the code for $N$ and so on. In practice  $|codeN| \leq log_2N +2log_2log_2N$ with two iterations and $|codeN| \leq log_2N +log_2log_2N+2 log_2log_2log_2N$  with three iterations and so on.  

If  $p_x$,  the probability of the occurrence of the `word' $x$ in a set of strings, is known, Shannon's noiseless coding theorem shows that an efficient self-delimiting code for $x$ can be chosen.  In which case, the code length for $x$ is  -$\lceil log_2p_x\rceil$ (i.e. $\log_2p_x$ rounded up).  Where all members of a set of $N$ strings occur with equal probability (i.e.  $1/N$) any string $x_i$ can be represented by a code of length $\lceil log_2N\rceil$.  This code is self-delimiting with respect to every other code in the set.  This integer is represented by $log_2N$ in the section on provisional entropy (used section  \ref {prov}) and following.  However, the decoding routine associated with the  code word for $x_i$  will usually require  information about the length the $code (x_i)$ to know when the code ends.  I.e. the routine must read each character in turn and decide whether it has read a sufficient number or not.   As a consequence an entropy measure will usually need to include $||code(x_i)||$ as well as $|code(x_i)|$.

The algorithmic entropy, measure by this process is an entropy measure of the exact state of the physical system at an instant of time.  Nevertheless, because the equilibrium state of a system belongs to the most probable set of outcomes, the expectation value of the algorithmic entropy, $\langle H_{algo}\rangle$, is virtually identical to the Shannon entropy ($H_s$), once allowance is made for differences of $O(1)$ \cite {6.,7.}. The significance of the algorithmic entropy concept is that a string representing a replicated structure has lower algorithmic entropy than other strings. For example, the algorithmic description of a crystal with its lattice structure is much simpler than a description specifying the position and momentum of each molecule in the liquid state.  This will be worked out in more detail in the following sections.

\subsection{Entropy relative to the common framework} \label {common}

If algorithms are self-delimiting, the Kraft inequality holds and separate optimal routines that are part of the main algorithm that defines the string can be concatenated (joined) with no ambiguity.  The entropy defined by combining maximally compressed routines is \(H_{algo}(s) = H_{algo}(t) + H_{algo}(s|t) +O(1)\).  The $O(1)$ refers to a simple instruction to link the routines \cite {8., 9.}. (Note this is to be distinguished from the quite different case where one programme might generate one string and another programme might generate an unrelated string.  In this case, as there may be no information in common and  $H_{algo}(s) \leq H_{algo}(t) + H_{algo}(s|t) +O(1)$).

Entropy is a state function and the entropy difference rather than the absolute entropy has physical significance in any physical situation.  As a consequence, the string representing common information or common instructions can be taken as given. This information \cite{10.} includes the binary specification of common instructions  equivalent to  ``$PRINT$'' and ``$FOR/NEXT$'' etc and the $O(1)$ uncertainty implied by the reference UTM.  For a physical systems such as a thermodynamic system, the algorithmic description of the physical laws embedded in the system and the algorithmic description of the graining of the phase space \cite {6.,11.,12.} are part of the common instruction string. In what follows the common instruction string will be represented by `$CI$' \cite{10.} and, given the common instructions, the physically significant entropy will be denoted by the conditional algorithmic entropy $H_{algo}(s|CI)$. Effectively a new zero of entropy is created by subtracting $H_{algo}(CI)$ from the full entropy.

\subsection {Provisional entropy} \label {prov}
A string that describes the state of a physical system often shows partial ordering but also shows noise and variation.  Such a sequence is recognized because there is an implicit reference to the pattern or the model that defines the criteria of set membership.  In other words, the set is recursively enumerable. Any sequence $s$ of this set $S$ can be at least partially compressed \cite{10.} as a string in the patterned set can be uniquely specified by an algorithm that identifies the particular string within the set together with an algorithm that defines the set.  Let $N_S$ be the number of members of the set $S$.  As all members are equally likely, as was shown in the previous section, a particular member can be identified by an algorithmic code of length $log_2N_s$, while the length of the code that defines the set can be denoted by $H_{algo}(S)$. This approach provides a ``best estimate'' of the entropy of a particular noisy patterned string representing the state of a physical system.  This entropy measure will be termed the provisional algorithmic entropy. The provisional entropy is derived by combining the the length of these two routines \cite{10.} giving; 
$$H_{prov}(s) = H_{algo}(S) +log_2N_s + O(1).$$
The second contribution is equivalent to the Shannon entropy of the set. In other words the provisional algorithmic entropy depends on the uncertainty in defining the member of the set together with the  algorithmic description of the pattern in the set. This is virtually the same as Zurek's ``physical entropy''\cite {6.}. Vereshchagin and Vit\'{a}nyi \cite {13.} derive the same result by finding the optimal model of set $S$ that contains the sequence and then the code for the sequence in the set. They call this description the Algorithmic Minimum Sufficient Statistic (AMSS) for a string. G\'acs \cite {11., 12.}, using an approach based on measure theory, derives a similar result.

While Devine \cite {10.} used the phrase ``revealed entropy'' to denote the entropy of a partially compressed pattern, here the term ``provisional entropy'' is preferred. The provisional algorithmic entropy is the best estimate of the length of the minimum description of the physical system given the available information.  The provisional entropy equals the true algorithmic entropy of a typical member of the patterned set.  However, a very few atypical members of the set may be further compressed.  One example is given by the provisional algorithmic entropy \cite{10.} of the noisy period 2 string, $1y1y1y1 \ldots 1y1y$ of length $N$ (where $y$ can be $0$ or $1$).  For this string, 
\begin{equation}
H_{prov}(s) \cong  N/2 + log_2(N/2) +  |1| + |0|. 
\end{equation}                 
Nevertheless the string $s' = 11111 \ldots 11$ from the same set has the much lower entropy; $H_{algo}(s')\cong log_2N +|1|$.  I.e. for this string, $H_{algo}(s')\leq H_{prov}(s)$.  Whenever this phenomenon is observed, it can be assumed the original model (i.e. the set capturing the pattern) needs to be revised. The approach is effective for any string that exhibits a noisy pattern that reflects data derived from a model.  As in the physical situation we will assume all the pattern has been recognised, the provisional entropy can be taken to be equal to the length of the description of the pattern or model and the term $log_2N_S$.  The latter term, corresponds to the closest integer to the Shannon measure of the uncertainty. As a consequence, one can slip between using entropy in a traditional sense where information about the set or configurations is taken for granted, and the algorithmic measure where this implied information is needed to more accurately define the system.

In what follows it is convenient to take the common information containing the physical laws etc as given. In other words the zero of entropy is taken to be the length of the algorithmic description over and above the given information.  Hence $H_{prov}(s)$ will be used to denote $H_{prov}(s|CI)$.  Choosing this zero of entropy highlights the fine structure contributions to the algorithmic entropy arising from terms such as $|0|$ or $|1|$.

\section{Replication}
Consider a classical physical system, initially isolated from the rest of the universe where the instantaneous microstate of the system is represented by a point in the multidimensional state space. Over time the state point of the system will move through state space under the operation of classical physical laws.  The connection between the trajectory of a system through its state space, and algorithmic information theory, can be made by recognizing that the instantaneous microstate of the system can be represented by a binary string (e.g. Zurek \cite {6.} and Li and Vit\'{a}nyi \cite {5.}). The size of the most concise algorithmic description of the configuration's microstate gives the algorithmic entropy.  Indeed, the algorithmic entropy of a system of replicates derived from an algorithm of the form: ``$REPEAT \: replicate \: description\: N \: times$'', will be low relative to a system of mainly dissimilar structures.  The algorithmic entropy measure now becomes key to inquiring into non- equilibrium processes.  

As the probability of replicates appearing increases with their occurrence, replication will drive the state point to a region of state space dominated by replicated structures.  In this case, each microstate in this region will be characterized by a large number of repeats of the replicate's description. As is discussed in more detail below, other physical processes, such as collisions or chemical reactions, will destroy replicates and the system will eventually settle in a region of long term stability.

\subsection{Algorithmic information theory and physical laws}
The process by which physical laws determine the trajectory of a classical physical system through its allowable states can be considered as a computation; the physical laws are algorithms embodied in a computer consisting of atoms, molecules or states.  In other words, the sequence of microstates that the system moves through under physical laws, maps on to a UTM that enumerates the microstates at each computational step.  As physical laws are reversible, the evolution of an isolated physical system will map on to a computing process that is logically reversible.  

However, while physical laws are reversible, a process in which energy and entropy are discarded is irreversible.  Such a process is described by an irreversible UTM.  The minimum programme that maps the actual physical evolution of the system, whether reversible or irreversible, will be denoted by `computation-from-within' and the length of the algorithmic description of the output `$o$' will be denoted by $K_{within}(o)$. However the algorithmic entropy $H_{algo}(s)$ implies that the algorithm describes the output from an external observer's point of view, rather than by the algorithm that maps the system's physical evolution. The difference between the two approaches is analogous to a flock of geese flying in a V-shaped formation.  The computation-from-within corresponds to each goose, adjusting (i.e. computing) its position and speed to minimize energy loss and maintain vision in such a way that the V-shaped trajectory emerges \cite {14.}.  Observer-computation on the other hand describes the observed V-shaped formation algorithmically.

Furthermore, $H_{algo}(o) \leq K_{within}(o)$, as the latter value arises from the programme the physical system is implementing.  This programme may specify irrelevant output from an external observer's point of view and therefore be a longer description.  Let $H_{algo}(o)$, the algorithmic entropy, be given by $|o^*|$, where $o^*$ is the minimal programme on the reference UTM to produce the output $o$. If $p^*(n)$ is the minimal programme that generates $o$ in $n$ steps by a computation-from-within, then $K_{within}(o)$ can be taken to be $=  | p^*(n)|$. As a consequence of the definitions $| p^*(n)|$ cannot be less than $|o^*|$. As the number of steps are unknown, $p^*(n)$ must contain a halt instruction when $o$ is reached; i.e. $p^*(n)$ must itself contain $o^*$.  In binary form the size of the programme string
\begin{center}
$|p^*(n)| \geq |o^*| + |increment \; instructions \; etc.|$.
\end{center}
As $p^*(n)$ size is dominated by $|o^*|$,  $K_{within}(o)  =  |p^*(n)| \geq |o^*|$.  I.e. $H_{algo}(o) \leq K_{within}(o)$. If $H_{algo}(o)$ is significantly less than $K_{within}(o)$ it indicates that the internal computation process takes some inefficient paths, or the process contains extraneous information not relevant to the externally observed pattern.  As a physical system is itself a UTM, this understanding helps to avoid confusion about when a computation is an entropy measure and when it is not.

\section{Entropy from reversible and irreversible algorithms}
The above example illustrates a key principle about reversible computation. At each computational step, information on the state of the system is stored in the `information bearing degrees of freedom' (IBDF) that are part of the system's algorithmic description. However, the process cannot be reversible if this stored information has to be discarded, as Landauer \cite {15.}, and subsequently Bennett \cite {7.,9.,16.} and Zurek \cite {6.} discuss. This occurs if entropy and information are lost to the system when an atom, energy or a photon is removed.  Discarding 1 bit of information from the IBDF contributes $k_BTlog_e2$ joules of energy to the environment.  Both logical and thermodynamic irreversibility correspond to the situation where information is removed from the computation. This removal leads to a reduction of entropy of the system, matched by an increase of entropy to the universe. 

Similarly entropy is gained by the system when information external to the system becomes part of it.  Generally it is convenient to encode this information as part of the input string that represents the physical or computational process. Zurek \cite{6.} and Bennett \cite{16.} have shown that the minimum entropy passed from the physical system to the universe in an irreversible process is $H_{algo}(i)- H_{algo}(o)$, noting that $H_{algo}(i)$ represents the information content of the initial state and $H_{algo}(o)$ that of the final state.  The change in entropy represents the difference between the minimal bits added to specify the original input string, and the minimal bits discarded in the final description of the output.  A nearly equivalent representation is $H_{algo}(i|o^*) - H_{algo}(o|i^*)$, where $i^*$ and $o^*$ are the compressed specifications of the input and the output. Zurek \cite{6.} argues that if one knew the exact description of these states, a cyclic process based on this information would be maximally efficient for extracting work. 
 
The reversible physical process generating a coherent photon outlined later in section \ref {coherent}, can map on to a logically reversible Turing machine that operates on the input string (representing the initial state) to produce an output string (representing the final state).  Examples of Turing computers based on physical or chemical processes are the ballistic computer \cite{18.}, and the Brownian computer \cite {7., 16.}.  Bennett illustrates the operation of the Brownian computer through the process of RNA polymerase copying a complement of the DNA string; reversibility is attained only at zero speeds. The process is driven forward by irreversible error correction routines that underpin natural DNA copying.  

In general, a Universal Turing Machine is not reversible, as more than one prior state can lead to a given state \cite {19.}.  This prior information is no longer accessible to the computing process unless it is recorded.  As discarding information about the prior state adds to the entropy of the external environment, the non reversible UTM describes a non reversible physical process. Nevertheless, a reversible process can be mapped on to a UTM by using a reversible algorithm that works in both the forward and the reverse direction, inserting the information that otherwise would be lost at the irreversible computational steps \cite {5.,16.,17.}.  Broadly speaking (see Landauer \cite {15.} for details) such a reversible must have the capability to:

\begin{itemize}
\item Generate the output from the input  
\item Copy the output
\item Reverse the computation to remove cumulated history and regenerate the input.  
\item Swap the regenerated input and the copied output to allow the reverse computation.
\end{itemize}

Let $KR(o|i)$ be the length of the reversible algorithm that maps the physical processes of a real world computation to produce output $o$ when given the input $i$.  However  $H_{algo}(o|i)$, the conditional algorithmic entropy of the output given the input, will in general be less that $KR(i|o)$ or its equivalent $KR(o|i)$ because of the likelihood that extra information must be included to make the process reversible. I.e.   $H_{algo}(o|i) \leq KR(i|o)$.

Thus the minimal reversible programme $p^*$ that generates the output from the input on the reference UTM may be shorter than the programme that maps the physical process.  For example, the programme implicit in the Brownian computer described by Bennett \cite{7.} may be able to be shortened if a path involving, say, a catalyst could be replaced by a more direct computing path.  The speed of such a process might be less, but the outcome would be the same.  Several authors \cite {20.,21.,22.} have considered the trade off between computer storage and number of computing steps to reproduce a given output.  The shorter the computation the more information storage is required to achieve the reversible computation.  For on going computations, the stored information must be erased and entropy is passed to the environment, making the process more difficult to reverse.

\section{Examples of replication}
The principles will initially be used to determine the provisional algorithmic entropy of a system of identical replicates as is outlined in sections \ref {coherent} and \ref {simple} below.  These simple idealised examples demonstrate a reversible physical process and how irreversibility implies the loss of information to the external environment. Section  \ref {var} following deals with the more complex situation where the replicates are not all identical but variation occurs.  Finally section \ref {birth} discusses the provisional algorithmic entropy for the situation where replicates are ``born'' and ``die'' in a far from equilibrium, homeostatic environment.

\subsection{Coherent photons as an example of a replicating system} \label {coherent}

A simple system that captures the essential nature of a replication process using the algorithmic information theory framework is that of stimulated emission.  A coherent photon is in effect a replicate of the stimulating photon.  With this in mind, let $X^0$ and $X^1$ represent the ground state and the excited state of an atom in a laser.  Consider a physical system made of $N$ such atoms together with the photons that are emitted when an excited state returns to its ground state.  The emitted photons can either be in an incoherent state, denoted by $P^x$, or in a coherent state denoted by $P^1$.  The state space of the system consists of;

\begin{itemize}
\item	the position and momentum states of the atoms;
\item	the position and momentum states of the photons;
\item the excited and ground state of each atom; and
\item the incoherent and coherent states of each photon.  
\end{itemize}

The instantaneous microstate of the system can then be represented by a point in the multidimensional state space.  Let the position-momentum space of the atoms and photons be divided into cells so that the instantaneous configuration of the system can be specified by placing a `$1$' in a cell corresponding to a particular particle's coordinates and a `$0$' for an unoccupied region.  If the number of cells is large relative to the number of particles, there will only be a scattering of 1's throughout this position-momentum space \cite {6.}. Where the state space spans the electronic states of an atom, a `$0$' denotes it is in the ground state and a `$1$' that it is in the excited state of the atom.  An $x,y$ or $z$ denotes the different incoherent state of a photon.  A `$1$' denotes its coherent state, as the coherent states can be assumed to differ only in their position coordinates.  At any instant of time a string of `$0$s' and `$1$s' and a mix of $x$'s, $y$'s and $z$'s defines the configuration of the system in this state space. In this example, the common information, which is taken as given, includes the physical laws as subroutines and the description of the coding process.

The length of the shortest algorithm that defines the microstate of the system represents its algorithmic entropy.  However initially it is convenient to discuss the replication algorithm symbolically rather than representing the state as a binary string.  Assume the system of photons and particles are contained within end mirrors and walls without energy loss.  As the $N$ atoms will be distributed randomly through the position and moment subspace, these degrees of freedom can be ignored. The initial state will be taken as the $N$ atoms existing in their excited states.  To illustrate the argument $N$ will be taken to be seven; i.e. the initial atomic state is $X^1X^1X^1X^1X^1X^1X^1$ and the photon states are empty.   This is an ordered state providing the high energy source for the replication process.  Sooner or later an incoherent photon $P^x$ will be emitted and the corresponding excited atom will return to its ground state denoted by $X^0$.  I.e.

\begin{center}
$X^1X^1X^1X^1X^1X^1X^1 \rightarrow P^x + X^1X^1X^1X^1X^1X^1X^0$;
\end{center}
noting that energy is conserved.  A coherent photon that arises through the stimulated emission of a photon from an excited atom is a replicate of the stimulating photon. The computation describing this replicating process can be represented by:
\begin {align}IF \; photon \; P^x is &\; within \; the \; stimulated \; emission \nonumber\\
&range \; of \; X^1 \;THEN \nonumber
\end {align}
\begin {center}
$P^xX^1X^1X^1X^1X^1X^1X^0 \rightarrow P^1P^1 X^0X^1X^1X^1X^1X^1X^0$,
\end{center}
where $P^1$, in contrast to $P^x$, represents a photon in a coherent state.  Over a period of time a number of coherent photons (denoted by $P^1P^1$ ....) will emerge.  
\begin {center}
$P^xX^1X^1X^1X^1X^1X^1X^0 \rightarrow P^1P^1 X^0X^1X^1X^1X^1X^1X^0$,
\end{center}
A possible outcome string is\\ 
$$P^1P^1P^1P^1P^1P^1P^1X^0X^0X^0X^0X^0X^0X^0.$$  
However, as material does not escape, each step of this replication process is reversible and a coherent photon may be absorbed, i.e. $P^1X^0 \leftrightarrows X^1$.   In practice, for most of the time, the system will settle in a stable configuration, where the birth and death of the coherent photon replicates will balance. The physical computation process is both thermodynamically and logically reversible.  There is no information loss or entropy change to the universe.    

It is now more convenient to specify the states of the system as a string rather than using the descriptive notation immediately above.  In which case the atomic states $X^1$ and $X^0$ are represented by a 1 and a 0 in the state space of the electronic states while $P^1$ and $P^x$ are represented by $1$ and $x$, $y$, etc. in the state space of the photon states. The position momentum states of the atoms can be ignored as they are random and unaffected by the computation process.  The input string, representing the a system of excited atoms with no photons, now takes the schematic form:
       
\begin{center}
   [Atom electronic states][Photon states] 
$i$=[1111111111111111111][empty states].\,\,\,\,\,\,\,
\end{center}
A representative outcome of the computation process is where some atoms are in the ground state and coherent (1)and incoherent (x,y) photons exist.  I.e.
\begin{center}
         [Atom electronic states][Photon states  ]
$o$=[0011111011011010110][111x1yz].\,\,\,\,\,\,\,\,\,\,\,\,\,\,\,\,\,\,\,
\end{center}
With these points in mind a general replicating process can be visualized as a computing process.  The input string $i$ specifies the components of the system, the nutrients and the energy source (the excited atoms). A programme string $p$ specifies the algorithm embodying the physical laws of the replication process which, in the above case, have not been made explicit.  The output string $o$ is generated by the computation on computer $U$ where $U(p,i) = o$. The initial state is more highly ordered that the output state `$o$' as the empty photon states are not needed for the original algorithmic description. The process is analogous to a free expansion where new states become available to the system, leading to an increase in entropy. While the final state is an equilibrium configuration, the replication process minimises the increase in entropy, ensuring the configuration is more ordered than it would otherwise have been.

\subsection{Replicating spins} \label {simple}
In contrast to the previous section where a free expansion occurred, this section deals with the more straightforward case.  A somewhat artificial model of a chain of random spins with four allowable orientations along the diagonals of a square will be used to further illustrate the principles of simple replication.  The particular model has been chosen because it is not so simple that it cannot be extended to the more general case, but it is sufficiently simple that the orientations of the spins can be coded in binary form in an intuitive way.  I.e. a typical sequence of these random spins is:
 
$$\searrow\nwarrow\swarrow \searrow \nearrow \nearrow \searrow \swarrow \nearrow \swarrow \searrow  \nearrow \searrow \nearrow \nwarrow \nearrow \searrow \nearrow \nearrow \searrow.$$
If the four different orientations are denoted by the coordinate of the arrow tip at $(\pm 1/2, \pm 1/2)$ the different orientations can be coded as (1,1), (1,0), (0,1) and (0,0) by adding $+ 1/2$ to the coordinate of the tip for each orientation.  Choose as the replicating spin the first spin and take its orientation as (10) corresponding to the direction $(1/2,-1/2)$.  Let the orientation of the remaining spins, be random and let each $x$ refer to a spin coordinate that is randomly a zero or a one. In which case the description can be coded as:

\begin{center}
$i = 10xxx \ldots xx \bf{000000} \ldots .$
\end{center}

The first two characters represent the replicate. The remaining random spins can only be aligned if thermal energy is transferred to an external sink that is more ordered. The string of $N$ zeros in bold type represents this initially ordered external sink.  It needs to be included as it is part of the computational system that is coupled to the spin system. Both the random section and the string of zeros will be taken to be of length $N$. Consider a replication process that aligns spins with the (1/2,-1/2) direction, i.e. the computation scans the string, when it finds the replicate  `$10$', it alters the next two characters to `$10$', transferring information to the ordered `$00 \ldots00$' section of the string.

In principle, the output string of aligned spins formed by a replication process is

\begin{center}
$o = 10101010 \ldots 10 \pmb{xxxxx \ldots xxxx}$.
\end{center}
The randomness has been transferred to what previously was the ordered section of the string (in bold type).  For large $N$, ignoring computing overheads, the information content of the input string is 

\begin{eqnarray}
K(i)& \approx N + log_2N \;(specifying \;the\; random\; 10xxxx..x)\nonumber \\ 
&+ log_2N  + |0| \;(from\; PRINT \; 0, \;N \;times ).       
\end{eqnarray}

The replication process can be specified by defining an operator $U_{10}$ that scans the string from the left hand end, two characters at a time and whenever a  `$10$' appears replaces the next two characters by a `$10$'. At each step, the ordered `$00$s' in the second section are replaced in turn by a reversible record of what has happened. This process is stepped through $N/2$ times. TABLE I gives the possible transformations at each step. 

\begin{table}[h]
\caption{The transition corresponding to the operation $U_{10}$.}
\begin{tabular} {@{}cccc@{}} \hline
Transition from & Transition to  \\ \hline
$xx \ldots 00$\hphantom{00} & \hphantom{0}$10 \ldots xx$ \\ 
$10 \ldots 00$\hphantom{00} & \hphantom{0}$10 \ldots 00$  \\
$11 \ldots 00$\hphantom{00} & \hphantom{0}$10 \ldots 01$  \\
$01 \ldots 00$\hphantom{00} & \hphantom{0}$10 \ldots 11$  \\
$00 \ldots 00$\hphantom{00} & \hphantom{0}$10 \ldots 10$  \\ \hline
\end{tabular} 
\end{table}

$U_{10}$ can be considered as an operator that rotates each spin in turn to align it with (1/2, -1/2) denoted by ($10$).  The corresponding ordered $00$ in the second part of the string is similarly rotated to keep a record of the each computation representing the transfer of the thermal excitation energy to a more ordered structure. 

The output after N/2 steps is:

\begin{equation}
(U_{10})^{N/2}i = 101010101010 \ldots 10 \pmb{xxxxxx} \ldots \pmb{xx}.                          
\end{equation}		

The information content of the output string is the length of the reversible programme that describes Eq. (7).  I.e.
\begin{align}
KR_{within}(o)  & \approx 	log_2N/2 + |10| \;\nonumber \\&\hphantom {00}(from ``PRINT \; 10 \; N/2 \; times") \nonumber\\
&+N+ log_2N  \;   (from \:the \; random \; string)\hphantom {000000000000} \nonumber \\
&+ | U_{10}| \; (specifies \; the \; replication \; process). \hphantom {000}                         
\end{align}
$KR_{within}(o|i)\approx 0$, i.e. $KR_{within}(o) \approx K(i)$ indicating that information has neither been created nor destroyed, as the information in $U_{10}$ the physical laws, is conserved. 
In general, an external observer will describe the system by the simple algorithm ``$PRINT \;10 \; N/2 \; times$'' and  
\begin{equation}
H_{algo}(o) \approx log_2N/2 + |10| + |PRINT|.                                     
\end{equation}
The difference between $KR_{within}(o)$ and the entropy measure $H_{algo}(o)$ shows that entropy algorithm does not record the string $\pmb{xxxx \ldots xx}$, which corresponds to the now random second section of the output.  Unless the $N$ bits in this random section are passed to the environment, the computation would remain reversible and the spins would not align in reasonable time frames. It is only when these $N$ disordered bits are ejected is the computation complete.  This ejection  increases the environmental entropy and corresponds to the latent heat cost of the ordering process ($k_BTln2$ joules per bit). 

When this is taken into account, for large $N$, and ignoring the contributions of $PRINT$ and $U_{10}$, the algorithmic entropy obtained by either an {\bf irreversible} computation from within, or by an external observer computation are virtually identical; i.e. 

\begin{equation} \label {nonvarying}
H_{algo}(o) \approx log_2N/2 + |10|.                                            
\end{equation}

\subsection{The entropy cost of replication with variations} \label {var}

The previous sections considered identical replicates.  However in physical situations replicates are not identical and this has an entropy cost. The algorithmic information theory approach is able to quantify the increase in entropy where variation occurs. A simple example, which can be generalised, illustrates the principle.  Variation can be included in the simple string $101010...10$ by allowing both `$10$' and `$11$' to be considered as valid variants of the replicate. This string then becomes the noisy period 2 string discussed in section \ref {common}, where every second digit is random. Thus a valid replicate system would now be the output string of length $N$ having the form $o = 1y1y1y1y..1y$, and where $y$ is $0$ or $1$.  There are $\mathcal{V} = 2^{N/2}$, members of the set of all possible replicate strings. The provisional algorithmic entropy, which is given in section \ref {common} includes the length of the code that identifies the string within the set, together with the specification of the pattern or the model that identifies the set itself. I.e. 

\begin{equation}
H_{prov}(o) \cong  |\mathcal{V}| + |description \; of \; pattern|.                       
\end{equation}
Here `$\cong$' is used to indicate that small inefficiencies in a specific algorithm can be ignored. As was mentioned above, to decode $|\mathcal {V}|$, the size of the specification for  $|\mathcal {V}|$ is needed.  I.e. $log_2N$, representing the length of the code for $N/2$, is needed. For the simple variants `$10$' and `$11$' the outcome is the same as that outlined in section \ref {prov} equation (5), i.e.

\begin{equation}
H_{prov}(o) \cong  N/2 + log_2N/2 +  |1| + |0|.           
\end{equation}
The provisional algorithmic entropy of a string with variation has increased by $N/2$ over the simple string of Eq. 9.

More generally let $r_1,r_2, r_3,r_\mathcal{M}$ be the $\mathcal{M}$ variations of the replicate $r$ of binary length $P$. Let $S_L$ be a mix of replicates of the form $S_L= r_ir_jr_jr_i \ldots $.  Each replicate $r_i$ can be represented by code($r_i$) and the string of codes can be compressed to form an overall code. In what follows we will assume no further pattern exists in the overall string; i.e. the mix of replicates is random.  The algorithm that generates the string must include the string's code and a decoding routine.  If the binary string is of overall length $N$, consisting of $L$ replicates existing in a variety of forms, then $L = N/P$, and there will be $\mathcal{V} = \mathcal{M}^L$ variations in the patterned set.  As there are $\mathcal M$ replicates, each replicate in $S_L$ can be coded with a self-delimiting code of length $log_2\mathcal M$. The length of the code for $S_L$ made up of $L$ replicates is $Llog_2\mathcal M$. In order to decode the code for $S_L$, each replicate code in the string must be fed into a subroutine that also specifies the length of the replicate, $P$, and the pattern in the replicate set. Thus the overall routine requires the input code of length $|\mathcal M^L|$, with $L$ and $P$  to be specified, noting that $N$ can be computed from  the product $LP$, and $|\mathcal M|$ from  $|\mathcal M^L|$.  Hence if ``$o$'' represents the string of variable replicators,

\begin{eqnarray} \label {provo}
H_{prov}(o) &\cong  Llog_2\mathcal{M} + log_2L  +  log_2P + |description \phantom {00}\nonumber \\
& of \; pattern \; of \; subset \; given \; P|  + O(1).                     
\end{eqnarray}
Here the $O(1)$ term includes minor computations such as the instruction that calls the subroutine to multiply $L$ and $P$.  The above assumed that there was no pattern in the way the replicates were arranged.  If each replicate $r$ is identical, further pattern exists, and in this case the $Llog_2\mathcal{M}$ term is unnecessary.  Equation (13) then becomes; 
\begin{equation}
H_{prov}(identical \; replicate \; string) \cong log_2L  + | r|.		        
\end{equation}
This is equivalent to equation (9) with $log_2L = log_2N/2$ and $|r| = |10|$.  For $N$ large relative to the pattern description,  $|r| \approx log_2P | + |description\;of \;pattern \;of \;subset|$.  Hence replicate variation increases the entropy by $Llog_2\mathcal{M}$.  In Shannon terms this is the increase in uncertainty due to the variation.

\subsection {Replication with variations and where replicates die and are born} \label {birth}
The previous examples considered a string of just replicates.  However in a situation of homeostasis, the replicates will reversibly coexist with non replicate structures.  The question are; ``What is the entropy cost of this?'' and ``How does the algorithmic entropy change?'' Again, the simple illustration provides general insights into complex replicating systems.  Consider the microstates of a system in an attractor-like region of its state space embracing a large number of replicated structures, coexisting with these non replicates. As the system trajectory moves from one microstate to another, replicates are reversibly created and destroyed. For example in the simple laser model described above, the laser line width indicates that there are variations in the specification of each coherent photon due to phase noise and fluctuations in the cavity size.  Furthermore, the coherent photons co exist with incoherent photons.

For simplicity it will be assumed that the replicate system has stabilized in a region of state space where there are a constant number $R$ of replicates of length $P_R$ and $X$ non replicates of length  $P_X$.  The configuration is represented by a string of length $N$. A typical string in the attractor region is of the form $S =r_1r_2x_1x_2x_3r_3x_5x_3x_3,...r_\mathcal{K}.$ Here $r_i$ represents the $i$th replicate from $M$ possibilities and $x_j$ represents the $j$th non replicate from $\mathcal K$ possibilities.

The Appendix shows that in this case the algorithmic entropy in the attractor region becomes:
\begin{align}
H_{prov}(S) & \cong  log_2\mathcal {P} + Rlog_2\mathcal{M} + Xlog_2\mathcal{K}  \nonumber \\
&+ log_2P_R  + log_2 P_X  + log_2R + log_2X \hphantom {00000000000000000000000}\nonumber \\
&+ |specification \:of  \:replicate\: pattern| \nonumber \\
&+ |specification \:of \:each \:non \:replicate|.
\end{align}

The shortest description will involve deriving one of $N$, $R$, $P_R$, $P_X$  and $X$ from the others (e.g. $N$). The term representing the Shannon entropy, the uncertainty measure is  $log_2\mathcal{P} + Rlog_2\mathcal{M} + Xlog_2\mathcal{K}$. This virtually the same as $log_2(number \; of \; members \; in\;set)$ \cite{10.}.  TABLE II shows how the entropy increases as the system becomes more complex. 

\subsection{The state space trajectory}
The approach can be used to inquire into the computational processes that determine the state space trajectory of a replicating system.  The following outlines a schematic representation of this evolutionary process.  Let the configuration be specified by a point in a state space consisting of replicates and non replicates.  The physical laws that determine the system's trajectory are embodied in structures of atoms and molecules that pass through the system. Let the computation-from-within process acting on the initial string $s$, be considered as a two stage computational step. At the first stage, the configuration changes when a computational operator $U_r$, representing the physical process of replication, on finding a replicate operates on a neighbouring non replicate and reproduces the replicate or a variation of it. At the second stage another operator $U_{phys}$, representing a non replicating physical process, increments the state space coordinates of each molecule or molecular component. For example, in the system of photons and atoms described above where photon coherence emerges, $U_{phys}$ shifts each atom's state coordinates by atomic collisions or by the absorption of a coherent photon. I.e. $U_{phys}$ operates on the given microstate of the system that in general will include replicated units.  

\begin{table}[t]
\caption{Increase of provisional algorithmic entropy with more complex systems of replicates and large $N$.}
\begin{tabular}{@{}cccc@{}} \hline
Entropy of fixed replicate  & $log_2L + log_2r$ \\
pattern         `abcd..e' &\\ \hline
&\\
Entropy of variable replicate  & $Llog_2\mathcal{M}+ log_2P + log_2L$\\
pattern with $\mathcal{M}$ variations & $+|pattern \:of \: replicates|$\\ \hline
&\\
Entropy of variable replicates &$log_2\mathcal{P} + Rlog_2\mathcal{M} + Xlog_2\mathcal{K} $   \\
with births and deaths & $log_2P_R +log_2P_X  + log_2R$ \\
&$ +log_2X +|pattern\: of\:replicates|$ \\
&$+ |non\:replicates|$ \\ \hline

\end{tabular}
\end{table}

After $t$ steps the outcome string is given by

$$(U_{phys}U_r)^ts =  s',$$
where $s'$ is an alternative microstate of the system.  While $U_r$ drives the computational process towards complete replication, $U_{phys}$ shifts the state point to a different coordinate of the allowable state space.  The trajectory through state space of a noisy replicating system takes place in three stages, as is illustrated in Figure 1.  The trends follow a discrete form of something like Eq. 1.

\begin{itemize}
\item	Initially, when no replicates exist, $U_{phys}U_r$ takes the microstate on a path that would at first sight pass through all the states in the system.
\item When a replicate appears as a segment of the string representing the current microstate of the system, $U_r$ duplicates the replicate.  $U_{phys}$, as part of the same computational step, will shift the microstate to its next configuration, perhaps modifying the replicate or dismantling it.  When there are few replicates their numbers will grow linearly with each step ${\rm d} n/ {\rm d}t \approx n/ \tau$     The trajectory of the system will trend towards an attractor region of state space where the microstates specify a large number of replicates.
\item	As the number of replicates increases, $U_{phys}$ will inevitably dismantle a replicate; essentially reducing the number in the microstate description.  For example, when the number of photons becomes high relative to the number of atoms in the ground state, a coherent photon will be absorbed and at a later time another will appear.  At this point, as the number of steps progresses, replicates will be destroyed and others will appear and the expectation value of d$n$/d$t$ = 0.  
\item	While there is no algorithmic information or entropy loss by the ejection of energy, or molecular species, the whole state space is accessible to the computing process, but the system will dwell for a long period in a region where the number of replicates is high.  However if there is a loss, as for example when latent heat is passed to the environment in a crystallisation process, the trajectory of the system will be constrained permanently in the attractor region of state space; the exact number of replicates fluctuating about a stable situation.  Provided that no further information or entropy  is exchanged, and all remaining steps in the computing process are reversible, the system will stay in the attractor region of state space. 
\end{itemize}

\subsection{Adaptation of replicates in an open system} \label {open}

The new resources flowing into an open system are seen as additions to the input string expanding its state space.  Similarly, resources flowing out of the system lead to a loss of information and a contraction of the state space. Where a changing input mix creates new computational paths available to $U_{phys}$ and $U_r$, some variations of the replicate may become less likely, while others may become more likely. The replicates that emerge will be selected (e.g. see Lifson et al. \cite {23.}) and will represent only a subset of possible replicates.  Ignoring the possibility that the system's trajectory will depend on the order in which certain variations (i.e. mutations) occur \cite {24.}, the attractor-like region of state space will become a smaller portion of the originally larger region.

The outcome is akin to Ashby's law of requisite variety \cite{25.}; to survive a system must generate sufficient variety to match the variety in the external environment.  When a different mix of information bits is added to the system as part of the input string, the real world computation will stall if the resources accessible to the physical system are not appropriate to carry the system to the next point of the trajectory. I.e. the system cannot generate sufficient variety to cope with change of the input states. Where sufficient variations of the replicate exist, the system will adapt, and the computation will settle in a narrow region of state space. 

\begin{figure}
\centerline{\includegraphics[width=15pc]{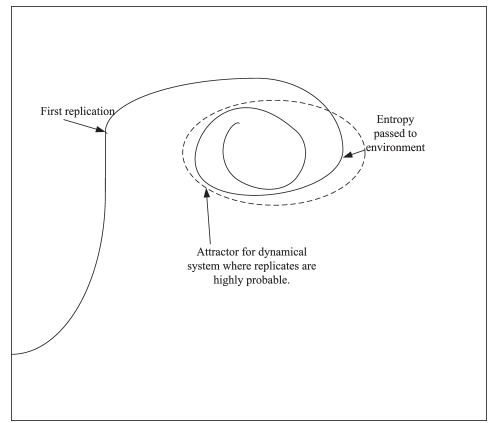}}
\caption {State space trajectory of replicating system.}
\end{figure}

\section{Irreversibility}

Replicates in an isolated attractor-like region of dynamical state space, die and are born as the state trajectory moves through the microstates of the subspace.  But if energy or material can diffuse elsewhere under the influence of the appropriate physical laws, the attractor-like \footnote{Strictly, as the state space variables are changing because material passes through, the region is best described as `attractor-like' rather than an `attractor'} region becomes far-from-equilibrium in terms of the whole universe of states.  In order to maintain homeostasis, it is necessary to replace the lost information and eject disorder, i.e. the waste material and heat that the system produces.
 
Removal of waste material and energy, or the addition of further inputs such as catalysts, may increase the probability of replication and therefore increase the rate at which the system settles in the homeostatic region of its state space.  If a system of, say, magnetic spins is to become more ordered, latent heat must be removed from the information degrees of freedom. Reversibility only occurs when the system is completely isolated, in which case, the net increase of replicates will be zero.  Forward drivers, such as error accumulation \cite{7.} or the removal of chemical material are needed, but these make the processes irreversible.

If the simple laser model is to be maintained off equilibrium, the lasing atoms will need to be pumped from an external source, e.g. by other photons. From a computational point of view, the information being passed to the environment from the information bearing degrees of freedom, must be compensated for by the information provided by the external photons that excite the lasing atoms. In this case, the attractor-like region maintains its basic shape but drifts through the hyper cube of the state space as similar degrees of freedom are added or removed.  If the structures in the input string that carry the physical laws governing the computational process are dismantled during the computational process, replacements are needed to allow the computation to continue. Other structures like catalysts exist throughout the computation and provide computing resources but remain unchanged as entropy flows in and out.  An account of all the entropy or information flowing in and out of a system implies that information is neither created nor destroyed, but is conserved.  While this does not seem to be completely consistent with the second law of thermodynamics, it is only at the level of the universe as a whole that information might appear conserved.  However, at this level, if a classical universe starts in a highly ordered state, the observer, as part of the universe has insufficient degrees of freedom to model the universe as a whole. As the algorithmic description of the universe must be from a reversible computing-from-within point of view its overall entropy increase is related to the number of steps undertaken in the computing process. Each step in the evolution of the universe as a whole may in principle be reversible.  The description of a state following the $t$ steps to the present time determines the algorithmic entropy of the universe (e.g. see \cite {6.}).  The algorithm describing the computation of the universe, which is the algorithmic equivalent of a free expansion to previously states that were not previously accessible, is something like,

\begin{eqnarray}
&STATE = initial\;state \nonumber \\
&FOR\;STEP= 0 \; to \; t \nonumber \\
&\hphantom{0000}\;\;Compute\;next\;STATE. \nonumber \\
&NEXT\;STEP\hphantom{00000}.
\end{eqnarray}
The length of the reversible algorithm is $KR_{within}(t) \approx |initial \; state| + log_2t  + |physical \; laws|$  and increases with $log_2t$.  As the universe is assumed closed, the most likely configurations are the completely disordered equilibrium states.  These will emerge after time $t'$ where, $log_2t' \gg log_2t$.  An external observer would describe the universe initially with a highly compressed algorithmic specifying the ordered state, but after a time $t'$ the universe would exist for long periods in configurations that are random or disordered.
 
\section{Coupling of replicator systems}
 
Replicates that use resources more efficiently will have survival advantages.  Autocatalytic sets of molecules, or forms of spatial clustering that become replicating units (e.g. DNA in a cell), settle in a more confined area of state space. Because resources are transferred within the set rather than escaping to the rest of the system, the overall loss of information is less and the replicating structure will be easier to maintain in an off-equilibrium region. When the output string from one replicating system is used as input information to another replicating system, both become interdependent and tend to stabilize each other; the off-equilibrium ordered state is maintained with a lower throughput of energy and information. For example, where photons from one laser system create a population inversion in another laser system, there is less entropy loss to the environment. The coupled systems co evolve by using resources more efficiently.  In a resource constrained environment, dependence will emerge in preference to alternatives.  As the coupled systems need less external resources they are more stable against input perturbations.  Their mutual attractor-like region will not drift through state space at the same rate as similar, but uncoupled, systems. 

\subsection{Nested systems}

Coupled systems can be nested to form a larger replicating system as, for example, when coupled cells are nested within functional tissue which in turn is nested in an organism.  Chaitin's \cite{26.} concept of `d-diameter complexity', which quantifies order at different levels of scale, applies to nested systems.

\begin{figure}
\centerline{\includegraphics[width=15pc]{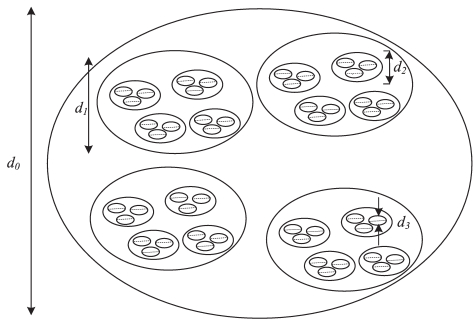}}
\caption {Nested structures at scales $d_0$, $d_1$, $d_2$ and $d_3$.}
\end{figure}

Figure 2 illustrates a nested system of replicators.  Let $H_{d0}$ represent the algorithmic entropy of the system, based on its minimal description at the largest scale $d_0$ (see Figure 3). Because nested structures can be described by nested algorithms, the algorithmic description is relatively short. However where the system is dismantled so that the large scale structure is destroyed and the system becomes a collection of structures at scale $d_1 < d_0$, the entropy increases as the large pattern is lost. The algorithmic description must specify each structure at this scale and how the structures are assembled.  For $d_1 \leq d \leq d_0$, $H_d > H_{d0}$, the entropy remains about the same until the scale reduces below that of $d_1$, the next level of pattern. The algorithmic description at the scale $d_2< d_1$ must include more detailed specifications and assembly instructions.  At the smallest scale, all the order is suppressed.  An exact detailed description of every piece of structure is required together with the description of the assembly algorithm.

\begin{figure}
\centerline{\includegraphics[width=18pc]{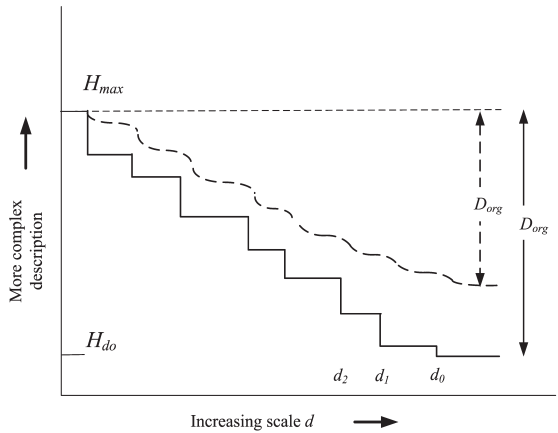}}
\caption {Variation of d-diameter complexity with scale; {\bf ---} nested replicators with no variation;   {\bf--\,--\,--\,--\,--}  nested replicators with variation; -\,-\,-\,-\,  No organization at any scale.}
\end{figure}

Figure 3 captures how the algorithmic entropy decreases as the scale is increased.  The stepped bold line shows an ideal system where the nested systems at each level of scale are identical.  As the scale of nesting increases, the entropy decreases.  The dashed line shows how variations in replicates at a given scale smooth out the steps in the ideal case, leading to a lower decrease of algorithmic entropy with scale.  However, where no organization at all exists the algorithmic entropy is the same at all levels of scale as shown by the dotted horizontal line, $H_{max}$ in figure 3.  Chaitin \cite{26.} quantifies the degree of organization ($D_{org}$) of structure $X$ by:

$$D_{org} = H_{max}(X) - H_{d0} (X).$$
The degree of organization, corresponds to Kolmogorov's  ``deficiency in randomness''.  This is a measure of how far a system is from equilibrium. If the previous discussion on an open system (section \ref {open}) is interpreted in terms of Figure 3, there is seen to be a trade off between high organization and high variety for a system to be viable.  Higher variety systems, with higher algorithmic entropy and therefore lower $D_{org}$ will be more viable off equilibrium, than  systems with high $D_{org}$, as the former have the flexibility to evolve within a larger region of their state space. As nesting can increase organization and thereby decrease entropy faster than any entropy increase due to variation, nesting compensates for the variation needed to ensure a structure is stable against change.  Indeed, this may be an inevitable consequence of selection processes acting on structures.  Interestingly, as software variation occurs at lower levels of scale, it would appear to be algorithmically more efficient to generate variation through software (e.g. variation in DNA) rather than directly.

\section{Conclusion}
This paper shows how algorithmic entropy provides a measure of the order in some simple models of replicating systems.  The models suggest that the process of replication may underpin much of the order observed in far from equilibrium systems.  Structures comprising order nested within order, characteristic of highly ordered living systems, can be analyzed within an algorithmic framework. Such nested structures may even be inevitable given the physical laws underpinning the universe, as nesting makes more efficient use of resources and partially compensates for variation in replicating processes.  While it is too difficult to provide a detailed algorithmic description of most real systems, the algorithmic entropy approach may well be useful in understanding incremental changes to real systems and, as well, provide broad descriptions of system behaviour.

\appendix 

\section {Replication with variation}

Let $S$ describe a microstate made up of a selection of $R$ replicates of length $P_R$ from the $M$ possibilities $r_1, r_2, r_3 ,...r_\mathcal{M}$, interspersed with $X$ non replicates of length $P_X$ from the $\mathcal K$ possibilities  $x_1,x_2 ,x_3 ,...x_\mathcal{K}$. The mixed string $S$ looks like $r_1r_2x_1x_2x_3r_3x_5x_3x_3,...r_\mathcal{K}$ and its length is given by $N = RP_R + XP_X$. The code for the string representing this configuration will consist of the code for each different replicate, with length $log_2\mathcal M$ and  the code for each non replicate with length $log_2\mathcal K$. However there will be many different arrangements  with the same replicates and non replicates. The number of these different arrangements in the string of replicates and non replicates can be determined if each replicate in $S$ is replaced by an `$r$'and each non replicate by an `$x$' where the subscripts are ignored in the first instance.  Let $S'$ represent this simplified string of a mixture $R$ of characters $r$ and $X$ of characters $x$. $S'$ has the form $rrxrr \ldots xxrx$. If there are $\mathcal P  = (R+X)/(X!R!)$ arrangements of $S'$ strings, each possible arrangement of replicates and non replicates can be represented by a code of length $log_2\mathcal P$. Once this code is determined, the detailed structure of each $r$ and $x$ can be specified by a subroutine. I.e. the algorithm that generates the string from the code can be obtained by combining the following separate algorithms, allowing for the O(1) instruction that concatenates the routines.
\begin {itemize}
\item An algorithmic routine of length $log_2\mathcal P$ that identifies the specific arrangement of replicates and non replicates.  The length of this routine can be denoted by  $H_{prov}(S')$
\item A routine that, given a particular arrangement, picks each replicate in turn and decodes it.  This routine can be denoted by \\$H_{algo}(replicates| S') = Rlog_2\mathcal{M} + log_2R  +  log_2P_R| \\+ |specification \:of \:replicate\: pattern|$.
\item A routine that given a particular arrangement picks each non replicate in turn and decodes it.  This routine is denoted by $H_{algo}(non \; replicate| S') = Xlog_2\mathcal K +log_2P_X + log_2X + |specification \:of \:each \:non \:replicate|$.
\end {itemize}

Putting these together 
\begin{eqnarray}
H_{prov}(S) & \leq H_{prov}(S') + H_{algo}(replicates| S') \nonumber \\
& + H_{algo}(non \; replicate| S') + O(1)\nonumber.
\end{eqnarray}

I.e. the algorithmic entropy in the attractor region then becomes:
\begin{align}
H_{prov}(S) & \cong  log_2\mathcal {P} + Rlog_2\mathcal{M} + Xlog_2\mathcal{K}  \nonumber \\
&+ log_2P_R  + log_2 P_X  + log_2R + log_2X \hphantom {00000000000000000000000}\nonumber \\
&+ |specification \:of  \:replicate\: pattern| \nonumber \\
&+ |specification \:of \:each \:non \:replicate|.
\end{align}


\begin{thebibliography}{}
\bibitem{1.} Szathmary, E. and Maynard Smith, J., From Replicators to Reproducers: the First Major Transitions Leading to Life, {\it Journal of Theoretical Biology} ({\bf187}, 555--571, (1997).

\bibitem{2.} Chaitin, G., On the length of programs for computing finite binary sequences, {\it J. ACM} {\bf 13}, 547--569, (1966).


\bibitem{3.}Kolmogorov, K., Three approaches to the quantitative definition of information, {\it Prob. Info. Trans.} {bf 1}, 1-7, (1965).

\bibitem{4.} Chaitin, G., A theory of program size formally identical to information theory, {\it Journal of the ACM} {\bf 22}, 329--340, (1975).

\bibitem{5.} Li, M. and Vit\'{a}nyi, P. M. B., {\it An introduction to Kolmogorov Complexity and its Applications, Second ed.} (Springer-Verlag, New York, 1997).

\bibitem{6.} Zurek, W. H., Algorithmic randomness and physical entropy, {\it Physical Review A} {\bf 40} (8), 4731--4751, (1989).

\bibitem{7.} Bennett, C. H., Thermodynamics of Computation- A review, {\it International Journal of Theoretical Physics} {\bf 21} (12), 905--940, (1982).

\bibitem{8.} Chaitin, G. J., Algorithmic Information Theory, {\it IBM Journal of Research and Developmen} {\bf 21}, 350--359,496, (1977).

\bibitem{9.}Bennett, C. H., Logical Depth and Physical Complexity, in {\it The Universal Turing Machine- a Half-Century Survey}, eds.~Herken, R. (Oxford University Press, Oxford, 1988), pp.~227--257.

 \bibitem{10.}Devine, S. D., The application of Algorithmic Information Theory to noisy patterned strings, {\it Complexity} {\bf 12} (2), 52--58 (2006)

\bibitem{11.} G\'{a}cs, P., The Boltzmann Entropy and Random Tests, http://www.cs.bu.edu/faculty \\/gacs/papers/ent-paper.pdf, (2004).

\bibitem{12.} G\'{a}cs, P., The Boltzmann Entropy and Randomness Tests- Extended Abstract, in {\it Proceedings of the Workshop on Physics and Computation} (IEEE Computer Society Press, 1994), pp.~209--216.

\bibitem{13.} Vereshchagin. N. K., and Vit\'{a}nyi, P. M. B., Kolmogorov's structure functions and model selection,
{\it IEEE Transactions on Information Theory}, {\bf 50}(12), 3265--3290, (2004) 

\bibitem{14.} Crutchfield, D., The calculi of emergence: Computational, dynamics, and induction, {\it Physica D} {\bf 75}, 11--54, (1994).

\bibitem{15.} Landauer, R., Irreversibility and heat generation in the computing process, {\it IBM Journal of Research and Development} {\bf 5}, 183--191, (1961).

\bibitem{16.} Bennett, C. H., Logical Reversibility of Computation, {\it IBM Journal of Research and Development 17}, 525--532, (1973).

\bibitem{17.}  Bennett, C. H., G\'{a}cs, P., Li, M., Vit\'{a}nyi, P. M. B., and Zurek, W. H., Information Distance, {\it IEEE Transactions on Information Theory} {\bf  44} (4), 1407--1423, (1998).

\bibitem{18.}  Fredkin, E. and Toffoli, T., Conservative logic, {\it International Journal of Theoretical Physics} {\bf 21}, 219--253, (1982).

\bibitem{19.} Bennett, C. H., Notes on Landauer's principle, reversible computation, and Maxwell's Demon http://xxx.lanl.gov\\/PS cache/physics/pdf/0210/0210005.pdf, (2003).

\bibitem{20.} Buhrman, H., Tromp, J., and Vit\'{a}nyi, P., Time and space bounds for reversible simulation, {\it Journal of Physics A: Mathematical and General} {\bf 34:35}, 6821--6830, (2001).

\bibitem{21.} Li, M. and Vit\'{a}nyi, P. M. B., Reversibility and adiabatic computation: trading time and space for energy, {\it Proceedings of the Royal Society of London}, {\bf Series A 452}, 769--789, (1996).

\bibitem{22.} Vit\'{a}nyi, P. M. B., Time Space and Energy in Reversible Computing, in {\it Proceedings of the 2005 ACM International Conference on Computing Frontiers}, Ischia, Italy, 2005, pp. 435--444.

\bibitem{23.}Lifson, S. and Lifson, H., A model of Prebiotic Replication: Survival of the Fittest versus Extinction of the Unfittest, {\it Journal of Theoretical Biology} {\bf 199}, 425-433, (1999)

\bibitem{24.} Yedid, G. and Bell, G., Macroevolution simulated with autonomously replicating computer programs, {\it Nature 420} (6197), 810-812, (2002).

\bibitem{25.}  Ashby, W. R., Introduction to Cybernetics (University Paperbacks, London, 1964).

\bibitem{26.} Chaitin, G. J., {\it Towards a mathematical definition of ``Life''}, in The Maximum Entropy Formalism, eds. Levine, R. D. and Tribus, M. (MIT Press, Massachusetts, 1979), pp. 477-498.


\end{thebibliography}
\end{document}